\documentclass[aps,prl,twocolumn,showpacs]{revtex4}
\usepackage{epsfig}
\usepackage{graphicx}

\begin{document}

\DeclareGraphicsExtensions{.eps,.EPS}

\title{Dipolar atomic spin ensembles in a double-well potential}
\author{A. de Paz$^{1,2}$, B. Naylor$^{1,2}$, J. Huckans$^{1,3}$, A. Carrance$^1$, O. Gorceix$^{1,2}$, E. Mar\'echal$^{1,2}$, P. Pedri$^{1,2}$, B. Laburthe-Tolra$^{1,2}$, L. Vernac$^{1,2}$}

\affiliation{$^{1}$\,Universit\'e Paris 13, Sorbonne Paris Cit\'e, Laboratoire de Physique des Lasers, F-93430, Villetaneuse, France\\
$^{2}$\,CNRS, UMR 7538, LPL, F-93430, Villetaneuse, France\\
$^{3}$\, Department of Physics and Engineering Technology, Bloomsburg University of Pennsylvania, Bloomsburg, PA 17815, USA}

%\newpage
\begin{abstract}
We experimentally study the spin dynamics of mesoscopic ensembles of ultracold magnetic spin-3 atoms located in two separated wells of an optical dipole trap. We use a radio-frequency sweep to selectively flip the spin of the atoms in one of the wells, which produces two separated spin domains of opposite polarization. We observe that these engineered spin domains are metastable with respect to the long-range magnetic dipolar interactions between the two ensembles. The absence of inter-cloud dipolar spin-exchange processes reveals a classical behavior, in contrast to previous results with atoms loaded in an optical lattice. When we merge the two subsystems,  we observe spin-exchange dynamics due to contact interactions which enable the first determination of the s-wave scattering length of $^{52}$Cr atoms in the S=0 molecular channel $a_0=13.5^{+11}_{-10.5}a_{B}$ (where $a_B$ is the Bohr radius).  \\

\end{abstract}
\date{\today}
\maketitle

\subsection{Introduction}

There has been tremendous activity recently to study physics associated with quantum magnetism using cold neutral atoms \cite{Quantum simulation, Quantum correlations}. Because of the very well controlled environment that can be achieved, experiments with cold atoms may even be used as quantum simulators, and pave the way to a better understanding of strongly correlated materials. Amongst the experimentally available systems based on cold atoms in optical lattices \cite{anderlini2007,trotzky2008,simon2011,meinert2013}, dipolar gases of magnetic atoms or polar molecules are interesting because dipole-dipole interactions provide a truly long-range coupling between spins, independent of tunneling-assisted super-exchange interactions \cite{barnett2006,micheli2006,gorshkov2011,peter2012,wall}. The long range interactions qualitatively change the physics of magnetism. For example in ferromagnets, dipole-dipole interactions play a crucial role to set the domain walls \cite{ashcroft} (see also the recent work with cold atoms in \cite{eto2014}). More generally, long range dipolar interactions may lead to new intriguing phases and anomalous behavior of spin systems \cite{peter2012}.

Despite the fact that recent experiments have observed spin-exchange dynamics of dipolar atoms or molecules in optical lattices \cite{3DSpinExchange,exchangeye}, the consequences of the nonlocal character of dipolar interactions on magnetism has not yet been neither explored nor visualized on a macroscopic scale. To explore this issue, we have loaded a dipolar chromium Bose-Einstein Condensate (BEC) inside a double-well trap. Our set-up produces two separated BECs which are close enough to each other that inter-well dipole-dipole interaction is not negligible on the experimental time-scale. We produce two spin domains of opposite polarization in the two wells, and describe their metastability with respect to dipolar interactions, a phenomenon well reproduced by a simple classical model describing two interacting giant spins.

The paper is organized as follows. First, we present a double-well trap for several thousand ultra-cold chromium atoms. This trap is created by the AC Stark shift of two interfering laser beams split by a lateral displacement beam splitter, and then guided by the same optics to the atoms. This insures common mode rejection and phase stability of the interference pattern. Second, we demonstrate a procedure to selectively flip the spin of the atoms in one of the two wells while atoms in the other well remain in their initial fully stretched spin state. We thus produce two clouds of typically 5000 atoms each, with opposite spin projections, whose mass centers are separated by 4.2 $\mu$m. The exchange part of the dipole-dipole interaction between the clouds is $\approx h\times 10$ Hz. We then monitor the spin populations as a function of the hold time, by means of a Stern Gerlach procedure. We observe no spin-exchange dynamics for timescales larger than 100 ms. This inhibition of the spin-exchange dynamics is in stark contrast with our previous results obtained with spinor chromium condensates featuring one or two spins pinned at lattice sites \cite{3DSpinExchange}.  We interpret this difference as due to the classical behavior of the two macroscopic ensembles of spins, each one behaving like a giant spin located in one well. Finally, we merge the two atom ensembles with opposite polarizations, and then observe spin-exchange dynamics due to spin dependent contact interactions. The analysis of our data provides the first determination of $a_{0}$ the chromium s-wave scattering length in the S=0 molecular channel.

\subsection{A double-well trap for condensed chromium atoms }

In this section, we present the experimental setup used in this work with special focus on the double-well potential that we superimpose on the Cr BEC. We use the most abundant isotope of chromium $^{52}$Cr. These atoms are bosonic particles carrying a spin S=3 and a large permanent magnetic dipole of 6 $\mu_B$ ($\mu_B$ is the Bohr magneton). Due to this large magnetic dipole moment, dipolar interactions between atoms impact Cr BECs properties \cite{lahaye2009, dipolarbec}. We have slightly modified our experimental setup described in \cite{CrBECus}: the crossed optical dipole trap wherein the BECs are produced now consists of two intersecting beams derived from a 100 W 1075 nm Yb fiber laser. The intensity of the laser beams is controlled by an acousto-optical modulator, whose radio frequency is modulated at 100 kHz.  This fast modulation creates a time-averaged potential, which is used to tune the anisotropy of the trap and optimize the 10 s ramp of evaporative cooling. At the end of evaporation, we produce a pure BEC with $10^{4}$ Cr atoms with trapping frequencies of $\omega_{x,y,z}=2\pi(520\pm12 ,615\pm15, 395\pm12 )Hz$ corresponding to Thomas-Fermi radii of $R_{x,y,z}=(2.5 ,2.1 , 3.3)\mu m$.

\begin{figure}
\centering
\includegraphics[width= 3.0 in]{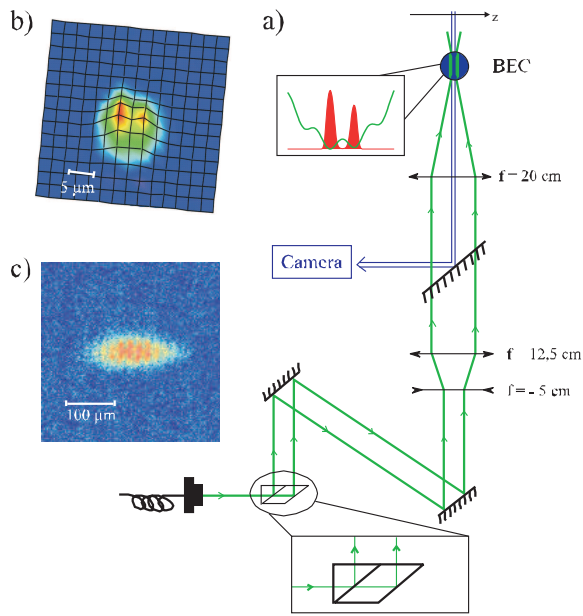}
\caption{\setlength{\baselineskip}{6pt} {\protect\scriptsize (Color online) a) Sketch of the double-well trap: a two-beam 532 nm laser interference pattern creates a spatially modulated trapping potential.  b) An average over 45 in-situ absorption images shows that atoms populate two wells. c) An absorption image taken after 5ms of time of flight reveals interferences between the two BECs. In these pictures, all atoms are in the $m_s=-3$ magnetic state.}}
\label{schema}
\end{figure}

We then proceed to loading the BEC into the double-well trap. This trap is an optical dipole trap which is created by the interference of two laser beams derived from a 532 nm single-mode cw laser. An incoming beam (with power up to 1.5 W) is split into two co-propagating parallel beams separated by 10 mm, using a non polarizing lateral displacement beam splitter (see Fig. \ref{schema}). These two beams are further separated by a 2.5 magnification telescope, and pass through a dichroic mirror with 90\% transmission. They are then focused onto the BEC using an achromatic doublet of focal length 200 mm. They interfere at the BEC location forming a 4.2 $\mu$m periodic potential along the $z$ direction. Given the 3 $\mu$m Thomas Fermi radius of the BEC along $z$, the system is therefore well suited to load the Cr atoms into two and only two minima of the trap. This realizes a double well trap. The separation between the center of the wells is 4.2 $\mu$m. The temperature of the beam splitter is actively stabilized to reduce thermal drifts of the interference fringes. Apart from the optical path in the beam splitter, both beams follow a parallel path through the same optics, which allows for common-mode rejection of the phase noise. This insures the stability of the double-well trap.

The dichroic mirror and the achromatic doublet are also used for absorption imaging. The optical setup is presented in Fig. \ref{schema} a). We also show in Fig. \ref{schema} b) an average of in-situ absorption images of the BEC in the double-well trap. Figure \ref{schema} c) is an absorption image after the BEC is abruptly released from the double-well trap, and a 5 ms free-fall time-of-flight. The two clouds interfere after being released from the trap \cite{andrews1997}. These pictures both show that the BEC is successfully transferred into a double-well trap, and that the transfer is smooth enough to produce two separate BECs in the two separated wells of the trap.

\begin{figure}
\centering
\includegraphics[width= 3.0 in]{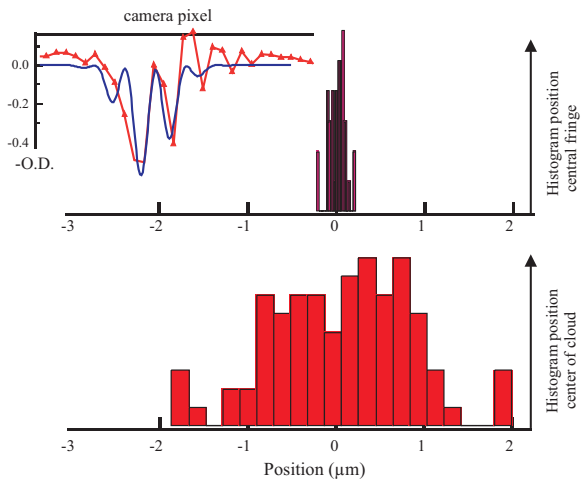}
\caption{\setlength{\baselineskip}{6pt} {\protect\scriptsize (Color online) Trap stability. The top curve shows (red triangles) the doubly integrated optical depth revealing the density distribution. The (blue) full line is a fit using a Gaussian modulated by a sinusoidal function. From the fitted phase of the sinusoidal function, we derive an histogram of the position of the central dark fringe, which is stable to 100 nm. From the center of the Gaussian, we derive histograms of the position of the atomic distribution, which reveals shot-to-shot noise in the BEC location of up to 3 $\mu$m.}}
\label{histogramme}
\end{figure}

We have also characterized the stability of the double-well trap by analyzing 45 in-situ absorption images, each corresponding to a different realization. As shown in Fig. \ref{histogramme}, the atom density is empirically fitted using a sinusoidal fit modulated by a gaussian. The noise on the fitted phase of the sinusoidal is a measurement of the phase stability of the double-well trap. The noise of the fitted center of the gaussian characterizes the position noise of the BEC. As seen from Fig. \ref{histogramme}, the phase stability of the double-well trap is excellent; however, the loading stability of the double-well trap is limited by the position stability of the BEC, which we attribute to the pointing noise of the IR beams.

We have measured the frequencies of the double-well trap by means of parametric excitation, for a total 532 nm light power of 380 mW. We apply an intensity modulation to the trap after the atoms are loaded, and we measure the heating as a function of the modulation frequency. We find within the experimental resolution the same trapping frequencies for both wells: $\omega'_{x,y,z}=2\pi(550\pm35, 700\pm50, 2835\pm325)$ Hz. The trapping frequency along $z$ can be boosted up to 6 kHz by increasing the green light intensity.

\subsection{Spin preparation}

In order to selectively spin flip the atoms in one of the two wells, leaving unchanged the spin of the atoms in the other well, we apply a radio-frequency ramp. To obtain the required selectivity, we apply a magnetic field gradient along the $z$ direction. The total magnetic field is practically parallel to $z$, with an amplitude $g\mu_B B_0/h=$ 200 kHz at the BEC position, and a magnetic field gradient $g\mu_B b/h$ of 2.5 kHz/$\mu$m along z. The strong confinement along the $z$ axis insures that atoms are pinned to the local minima of the trap, and that their motion is approximately insensitive to the magnetic field gradient.

After the magnetic field gradient is applied, we perform an rf frequency sweep, whose intensity follows a Gaussian temporal profile. Such pulse shaping (similar to the Blackman window commonly used in atom interferometry \cite{kasevich1991}) is necessary to avoid the fast frequency components associated with instantaneous turn-on and turn-off of the rf field, which in practice are sufficient to spoil the selectivity of the rf sweep. With a 5 ms rf sweep, whose span and peak Rabi frequency are 30 kHz and 1.5 kHz respectively, we successfully flip the atomic spins selectively in one well, with an efficiency close to 90\%. We then switch off the magnetic gradient, to enable intersite spin-exchange \cite{3DSpinExchange}.

To measure the evolution of the different spin populations, we use a Stern-Gerlach procedure, which separates the different spin states with a magnetic field gradient (applied after a given hold time $t$ during which spin dynamics may take place). In the absorption image shown in Fig. \ref{stability} (taken for $t=0$), atoms on the left correspond to negative $m_s$ states, whereas atoms on the right correspond to positive $m_s$ states. The signal asymmetry is due to different efficiencies in the absorption imaging of the different $m_s$ states.

The Stern-Gerlach image shown in Fig. \ref{stability} was taken after an rf-sweep which only addressed the atoms in one of the wells. Our interpretation of this image is the following. The atoms at the left of the picture are atoms unaltered by the rf. They were at the left-well position, and they remained in $m_s=-3$. Atoms at the right of the picture have been affected by the rf, because they are in the right well. The rf sweep has promoted them to $m_S=3$. As a consequence, this image provides a determination of the state of the atoms independently in the left well and the right well. Following the system preparation spin mixing mechanisms can eventually populate all spin states. Nevertheless, for short time evolution, positive $m_s$ states originate from the well where the spin flip was efficient (the right well), while negative $m_s$ states originate from the left well. The Stern Gerlach measurement will therefore allow us to study the spin dynamics of each well separately as long as the population in $m_s=0$ remains negligible. A site-selective Stern-Gerlach detection would only be necessary at longer times.

\subsection{Metastability with respect to inter-site spin-exchange}

We first discuss the spontaneous evolution of the spin distribution after the right atoms are promoted to the $m_s=3$ state, and the magnetic field gradient is switched off.

\begin{figure}
\centering
\includegraphics[width= 3.0 in]{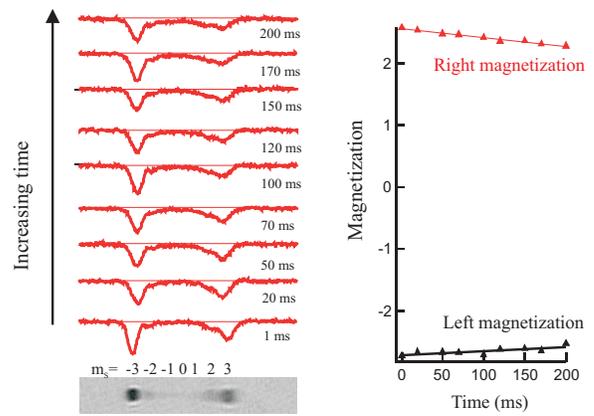}
\caption{\setlength{\baselineskip}{6pt} {\protect\scriptsize (Color online) Spin dynamics in the double-well trap. Lower left panel: Absorption image after the Stern Gerlach separation demonstrating the spin preparation with opposite polarization in the two wells. Top left panel: density profiles for increasing hold time before release and Stern Gerlach analysis showing the evolution of the spin composition. Spin dynamics is very slow, and the magnetization is almost constant in both wells (right panel).
}}
\label{stability}
\end{figure}

Initial spin dynamics in a fully separated double-well trap is purely dipolar, because atoms are locally in a stretched state (see below). For an initial state $\left|N:L,-3;N:R,3\right\rangle$ (corresponding to $N$ atoms in the left well in state $m_s=-3$ and $N$ atoms in the right well in state $m_s=3$), two spin relaxation channels are possible. The first one is local and corresponds to dipolar relaxation induced by collisions between $m_s=3$ atoms inside the right well. We do observe such phenomena (see Fig. \ref{stability}). Dipolar relaxation is in fact extremely rapid and we cannot resolve its dynamics. Indeed, the density of the atoms in the double-well trap is extremely high (close to $10^{21}$ m$^{-3}$), which, at the magnetic fields at which we operate, results in sub-ms dynamics. However, because the energy released in a dipolar relaxation event is smaller than the trap depth, atoms remain in the right well after dipolar relaxation. The inelastic process thus results in a rapid increase of the temperature of the right cloud, up to 3 $\mu$K. The density therefore rapidly decreases, to typically 4.10$^{19}$ m$^{-3}$, so that dipolar relaxation practically stops after 1 ms : given the rate parameter of dipolar relaxation at the experimentally applied magnetic field (typically 50 mG)\cite{pasquiou2010}, the dipolar relaxation rate is then estimated to be on the order of 5 $s^{-1}$ . As shown in Fig. \ref{stability}, the right magnetization then barely decreases for times up to 200 ms.

The second process which can occur is dipolar spin-exchange between the right and the left atoms. The spin-exchange rate between (2N=10$^4$) chromium atoms separated by a distance of $d=4.2$ microns is $\hbar \Gamma_{exc}= S N\frac{\mu_0}{4 \pi} \frac{(g_s \mu_B)^2}{d^3}$ (with $\mu_0$ the magnetic constant, and $g_s \approx 2$ the Land\'e factor for ground state $S=3$ Cr atoms). We find $\Gamma_{exc} =2\pi \times 10$ Hz. As shown in Fig. \ref{stability}, we do not observe spin-exchange, even for interaction times much larger than $1/\Gamma_{exc}$: after the initial fast dipolar relaxation, the populations remain almost frozen in their spin states. This metastability is one of the main results of this paper.

The absence of spin dynamics which is shown in Fig. \ref{stability} is in stark contrast with our previous measurements made for an array of individual atoms in an optical lattice \cite{3DSpinExchange}. In this latter case, the spin-exchange rate was measured to be in good agreement with many-body theory, and in qualitative agreement with the rate due to dipolar exchange interactions  between two atoms separated by the lattice spatial periodicity \cite{3DSpinExchange}. We show in the next paragraphs that the metastability of the spin distribution observed here is a signature of the difference between quantum magnetism \cite{3DSpinExchange} and classical magnetism shown in this work.

\subsection{Interpretation of spin-exchange suppression}

In the present experiment, atoms are initially prepared in both wells in a stretched state, therefore spin dynamics associated with contact interactions is gauged out if the two wells are fully separated. Two atoms locally interact only through the $S=6$ molecular potential, and no local spin-exchange is possible \cite{dsk2013}.

Furthermore, the interaction of a pair of atoms in the right well with atoms in the left well, through dipolar interactions, does not modify the molecular channel in which this pair of right atoms interact. Physically, this can be understood by recalling that dipole interactions consist of the interaction of one atom with the magnetic field created by another atom. Therefore, the right pair of atoms simply undergoes Larmor precession around the field created by the left atoms, leaving the molecular potential unchanged. As a consequence, atoms in a given well interact at all times through the $S=6$ molecular potential, i.e. in an eigenstate of contact interactions for which contact spin-exchange is ruled out.

We therefore developed a theory to account for the dynamics observed, where the spins of the left well interact only with the spins of the right well (and vice-versa) through dipolar interactions. Within the Heisenberg picture the equation of motion (for left well spins) reads:
\begin{equation}
\frac{d}{dt}\hat {\vec s}_{L,i}=\gamma \hat {\vec s}_{L,i}\wedge \left(\vec B_0+\sum_j\vec B(\hat {\vec s}_{R,j}, \hat {\vec r}_{i,j})  \right)
\end{equation}
with $\gamma=g_s \mu_B/\hbar$ the gyromagnetic factor, and analogously for the spins of the right well.
The magnetic field generated by a single spin at a position $\vec r$ from the spin location is:
\begin{equation}
\vec B(\hat {\vec s} , \hat {\vec r})=\frac{\mu_0 \gamma }{4\pi} \frac{3\hat {\vec r}\: (\hat {\vec r} \cdot \hat {\vec s})-\hat {\vec s}\: \hat{r}^2}{\hat{r}^5}
\end{equation}

In order to simplify the treatment of the problem,  we define the total spin of the left well:
\begin{equation}
\hat {\vec S}_L= \sum_i \hat {\vec s}_{L,i}
\end{equation}
and analogously for the right well ($S_L=S_R=Ns$). We finally obtain two simple equations of the following form:
\begin{equation}
\frac{d}{dt}\hat {\vec S}_{L}=\gamma \hat {\vec S}_{L}\wedge \left( B_0 \vec u_z+\vec B(\hat {\vec S}_{R}, \vec d)  \right)
\label{macrospin}
\end{equation}
$\vec d$ being the relative position of the two wells ($\langle\vec r_{i,j}\rangle\simeq \vec d$).

The second term of eq.(\ref{macrospin}) can lead to evolutions for $\left<\hat{S}_{L,R\:z}\right>$. At high external magnetic field, dipolar relaxation is energetically forbidden in our theoretical framework, which does not include the mechanical degrees of freedom. Then, the effective Hamiltonian only comprises Ising and spin-exchange terms. As a consequence, the following effective Hamiltonian gives the evolution of the system:
\begin{equation}
\!\! \hat{H}_d^{eff}\! =\!  -2\frac{\mu_0 \gamma^2}{4\pi d^3}\! \left(\hat{S}_{L}^{z}\hat{S}_{R}^{z} \! - \! \frac{1}{4}\left(\hat{S}_L ^+ \hat{S}_R ^- \! + \!  \hat{S}_L ^- \hat{S}_R ^+\right)
\! \right)
\label{HdSE}
\end{equation}
This effective Hamiltonian drives as well the spin evolution of dipolar atoms in optical lattices \cite{3DSpinExchange} when suppression of dipolar relaxation is obtained, for an external magnetic field smaller than the smallest band gap of the lattice in that case.

\begin{figure}
\centering
\includegraphics[width= 3.0 in]{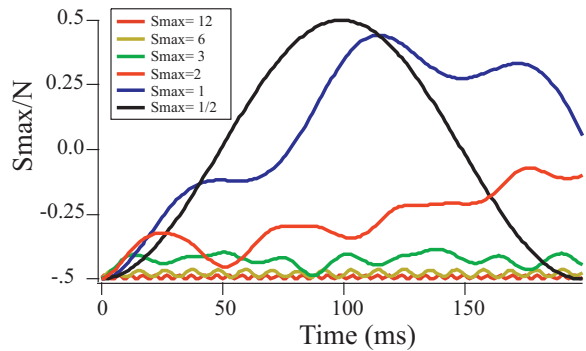}
\caption{\setlength{\baselineskip}{6pt} {\protect\scriptsize (Color online) Magnetization dynamics in one well for increasing number of particles. Here we consider spin 1/2 particles; the increasing number of atoms per well create an increasing local total spin Smax (from 1/2 for one atom per well, to 12 for 24  particles per well). Magnetization per atom in a given well is plotted as a function of time. Increasing spin results in almost frozen spin dynamics, therefore reaching a classical behavior.}}
\label{increasen}
\end{figure}

Equation (\ref{macrospin}) readily shows that the many-body evolution of 2$N$ spins in two traps can be reduced to the two body evolution of two giant spins. It can be numerically solved, and the time evolution of the left magnetization as a function of $N$ is shown in Fig. \ref{increasen} for the case where the external magnetic field dominates the dipolar field created by the atoms of the right well, in which case eqs.(\ref{macrospin}) and (\ref{HdSE}) give identical results. This condition applies to our experiment as typical values of these fields are respectively 20 mG and  30 $\mu$G. This analysis shows that spin-exchange due to dipole-dipole interactions between the two wells is strongly inhibited when the magnitude of the spin increases, which confirms our experimental data.

Our experimental data is also well accounted for by a classical theory of two localized interacting magnets of opposite magnetization in a large external magnetic field. In this case, magnetization dynamics is also completely frozen. Indeed, provided intersite spin correlations are neglected, eq.(\ref{macrospin}) reduces to the simple equation of precession of classical magnets (also used in the framework of spintronics for example \cite{levybook}). Within this classical approximation, and in the presence of an external magnetic field larger than the field created by the magnets, the orientations of two classical magnetic moments of opposite directions, each almost parallel to the magnetic field, are locked. This results from dynamical stability of a system which is otherwise energetically unstable. Our experimental observations can therefore be understood within this relatively simple classical magnetism framework.

A closer look at the quantum model shows that the spin dynamics of $N$ spins in fact does not completely vanish at short times. We numerically observe spin-exchange collisions, where two atoms undergo the transition $(-3,3)\rightarrow (-2,2)$, within a timescale $1/\Gamma_{exc}$. However, these collisions do not proliferate, and the spins remain roughly locked at their initial positions for extremely long durations. The reason why massive spin-exchange cannot occur is that the Ising term $\hat{S}_L^z\hat{S}_R^z$ of eq.(\ref{HdSE}), creates an energy barrier which the exchange terms $\hat{S}_L^+\hat{S}_R^-+\hat{S}_L^-\hat{S}_R^+$ are too small to overcome for large spins. As the size of the total spin increases, the spin fluctuations due to spin-exchange processes decrease relative to the total spin. This phenomenon is reminiscent of the $1/N$ paradigm in solid state physics, which states that spin fluctuations decrease as $1/N$ for a composite spin of length $N$ made of $2N$ electrons, because the fluctuations only involve the tunnelling of one spin excitation ($s=$1/2 in the case of an electron) (see for example \cite{wu2010}). As $N$ increases, we can observe in the numerical simulations the cross-over between quantum magnetism and classical magnetism.

Equation (\ref{macrospin}) also predicts that non classical spin states can be produced at timescales much longer than typical available experimental times. In particular, a quantum tunnelling to a state close to $\left|N:R,-3;N:L,3\right\rangle$ is expected at a well defined time $T_{\mathrm{t}}$. Interestingly, at $T_{\mathrm{t}}/2$, the state of the system is highly entangled, close to $\frac{1}{\sqrt{2}}(\left|N:R,-3;N:L,3\right\rangle -i \left|N:R,3;N:L,-3\right\rangle)$. We are now studying a generalization of this formalism using radio-frequency pulses, which may enable the creation of a Schrodinger cat state in a timescale $1/\Gamma_{exc}$ (per atom).

\subsection{Observation of spin-exchange due to contact interactions}

A closer inspection of the data shown in Fig. \ref{stability} reveals a very slow and small increase in the $m_s=-2$ population. Such an increase is inconsistent with the physical picture introduced in the previous paragraphs, wherein spin-exchange results from inter-site dipolar interactions. To understand this non-vanishing spin dynamics, we have further characterized our trap parameters.

First, from the measured trapping frequency along the $z$ direction, and from the measured periodicity of the optical lattice applied to the atoms, it is straightforward to deduce the height of the barrier separating both traps. We find that the barrier has a maximum height of 200 kHz for a maximal optical power of 1.5W. To estimate whether both clouds are indeed well separated, it is important to carefully measure the energy distribution of the atoms within the trap. As stated above, before the spin preparation is performed, the atoms remain in the quantum degenerate regime, and both clouds are indeed very well separated, the 350 nm Thomas Fermi radius being much smaller than the 4.2 $\mu$m distance between the traps. As described before, dipolar relaxation in the $m_s=3$ state is very fast. Because the energy released in dipolar relaxation is smaller than the trap depth, dipolar relaxation results in a strong heating of the cloud wherein $m_s=3$ states are produced. Absorption pictures after the atoms are released from the trap, in the presence of a magnetic field gradient to separate the (left) $m_s=-3$ atoms from the (right $m_s=3$ atoms) indeed reveal that the temperature at the right rapidly increases to typically 3 $\mu$K (roughly twice the BEC transition temperature for 5000 atoms within one well of this trap). Although such temperature corresponds to a typical energy of 60 kHz, smaller than the energy barrier, a Boltzmann distribution implies that about 20 percent of the atoms have enough energy to surpass the energy barrier between the traps.

We therefore interpret the slow spin-exchange dynamics shown in Fig. \ref{stability} as resulting from contact-interaction-driven spin-exchange between left well $m_s=-3$ atoms, and atoms from the right well, mostly in the $m_s=3$ state, which have enough kinetic energy to overcome the barrier and collide with the right atoms.

\subsection{A determination of the $S=0$ scattering length}

We have further investigated this scenario by adiabatically removing the optical barrier between the traps, after spin preparation is performed. The barrier is lowered in 20 ms, and the magnetic field gradient is turned off. After this procedure, the cloud consists of a mixture of mostly $m_s=-3$ and $m_s=3$ atoms. The measured temperature is 1.2 $\mu$K, corresponding to 2.5 $T_c$, $T_c$ being the critical temperature for Bose-Einstein condensation. The trap, characterized by parametric heating spectroscopy, has trapping frequencies of $\nu_{x,y,z}=(520, 615, 400)$ Hz. Interestingly, the peak density of $m_s$=3 atoms is then $3.5 \times 10^{18}$ m$^{-3}$, which insures that dipolar relaxation can be neglected for timescales below 1s. Indeed, we then observe a spin dynamics which is, to within our signal to noise ratio, at constant total magnetization for the first 500 ms.

\begin{figure}
\centering
\includegraphics[width= 3.0 in]{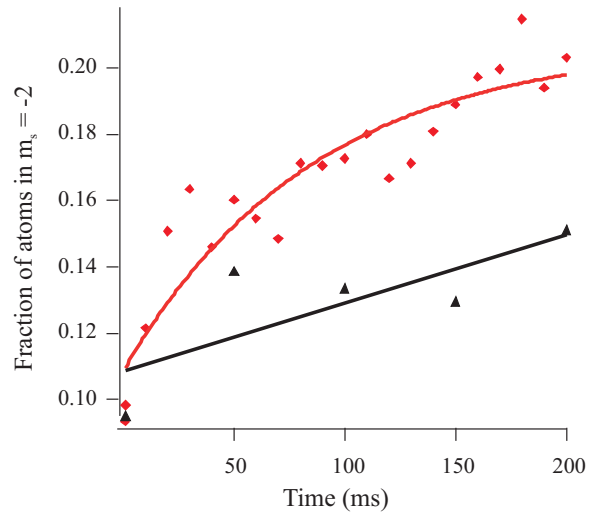}
\caption{\setlength{\baselineskip}{6pt} {\protect\scriptsize (Color online) Comparison of the spin dynamics with or without recombination. Population in $m_s=-2$ is plotted as a function of the hold time. Red diamonds: in the recombined trap; black triangles: the barrier between wells is maintained during spin dynamics. Each point is the result of an average of typically 8 measurements. Full lines are results of fits to guide the eye.
}}
\label{dynamics}
\end{figure}

\begin{figure}
\centering
\includegraphics[width= 3.0 in]{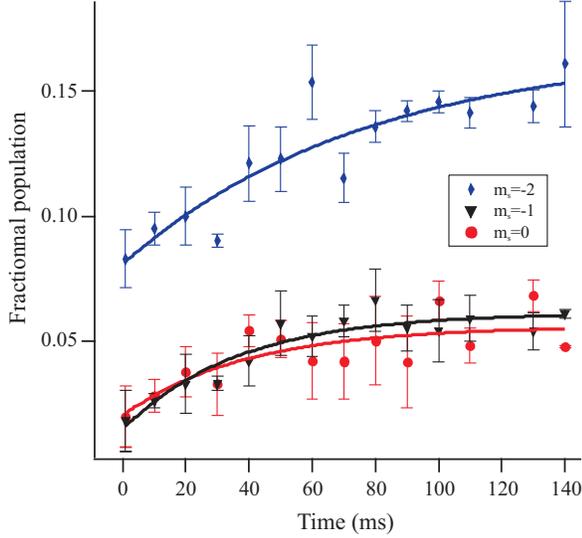}
\caption{\setlength{\baselineskip}{6pt} {\protect\scriptsize (Color online) Evolution of the spin composition after recombination. Fractional populations in $m_s=-2,-1,0$ are plotted as a function of the hold time in the recombined trap. Error bars show statistical uncertainties. Full lines show exponential fits to guide the eye.
}}
\label{dynamics2}
\end{figure}

We have focused on the spin dynamics in the recombined trap for the first 200 ms after recombination. Experimental data is shown in Figures \ref{dynamics} and \ref{dynamics2}. We observe a decrease of the $m_s=-3$ population, and an increase of $m_s=-2,-1,0$. In Fig. \ref{dynamics}, we compare the evolution of population in $m_s=-2$  for the recombined trap to the spin dynamics which is observed when the traps are not recombined. The timescale of spin dynamics when the two wells are recombined is faster than the timescale when atoms are kept separated. Although both timescales are not vastly different, it is useful to stress that the peak density in the double-well trap is $4 \times 10^{19}$ cm$^{-3}$, significantly higher than the peak total density in the recombined trap, $7 \times 10^{18}$ cm$^{-3}$. Two consequences can be drawn from these observations. First, spin dynamics relies on a good overlap between the two $m_s=-3$ and $m_s=3$ clouds. Second, as the spin dynamics in the double-well is typically twice slower despite 6 times stronger local densities, the overlap between $m_s=-3$ and $m_s=3$ clouds in the double well is necessarily very low. This confirms that these two clouds remain very well separated before we lower the barrier: we estimate that about 500 atoms may have moved from one well to the next to be consistent with observed spin-dynamics timescale when the barrier is not removed. The fact that both spin states are still physically well separated is crucial for our interpretation of the spin-exchange suppression within the double-well lattice in terms of classical magnetism.

To account for the observed spin dynamics in the merged traps, endoenergetic magnetization changing collisions (for example leading to demagnetization cooling in \cite{fattori2006}) are ruled out, because the magnetic field is much larger than the temperature of the cloud, which results in a negligible population of $m_s=-2$ at thermal equilibrium. The spin dynamics therefore results from spin-exchange interactions, which can be either triggered by dipolar interactions of by spin-dependent contact interactions.

As a first study, we have fitted the time evolution of $m_s=-2$ as a function of time, to deduce a scattering cross-section. We assume that the time evolution of a spin population is set by the following equation:

\begin{equation}
\frac{dn_{-2}}{dt}= \beta n_{-3} n_{3}
\label{dyn}
\end{equation}with $\beta = \sigma \bar{v}$ a rate constant, $\sigma$ the cross-section, $\bar{v}=4 \sqrt{\frac{k_B T}{\pi m}}$ the average atomic relative velocity, and $n_i$ the density in state $m_s=i$. We assume the temperature of $m_s=-3$ and $m_s=3$ atoms to be identical. This is realistic for times larger than 10 ms after the merging of both clouds, given the temperature and density of the cloud, and the typical scattering length $\approx$ 100 $a_B$, which leads to a typical elastic collision rate of 200 s$^{-1}$. Integrating eq.(\ref{dyn}) over space yields:
\begin{equation}
\frac{dN_{-2}}{dt}= \frac{n^0_3 \sigma \bar{v}}{2 \sqrt{2}} N_{-3}
\label{dynInt}
\end{equation}
with $N_{m_s}$ the number of atoms in state $m_s$. Thus the time constant for the time of evolution of the $m_s=-2$ population, relative to the $m_s=-3$ population, is given by
\begin{equation}
\frac{1}{\tau}=\frac{n^0_3 \sigma \bar{v}}{2 \sqrt{2}}
\label{tau}
\end{equation}
where $n^0_3$ is the peak density of $m_s=3$ atoms. The largest uncertainty for the experimental determination of $\sigma$ is $n^0_3$ is due to fluctuations in the spin preparation (associated with position fluctuations of the IR beam, see Fig. \ref{stability}). To analyse our experimental data we use a linear fit based on the first 50 ms dynamics, which yields $\sigma_{exp ,(-3,3)\rightarrow (-2,2) } = (1.25 \pm 0.5\pm0.4) \times 10^{-17} $m$^2$, where we indicate successively statistical and systematic errors.

We first compare such experimental cross-section to the predicted cross-section for spin-exchange interactions due to dipolar interactions. Within the Born approximation, we estimate  $\sigma_{dip} = 1.4 \times 10^{-18} $m$^2$ \cite{pasquiou2010}. Such cross-section is insufficient to account for the observed spin dynamics, which we therefore attribute to spin-exchange associated to spin-dependent contact interactions.

Due to the spin-dependency of contact interactions, three-different spin-exchange mechanisms are allowed from the initial mixture of states $(-3,3)$. These channels are (i) $(-3,3)\rightarrow (-2,2)$,(ii) $(-3,3)\rightarrow (-1,1)$ and (iii) $(-3,3)\rightarrow (0,0)$. Within the Born approximation, the scattering cross-sections for each of these individual channels are:

\begin{eqnarray}
\sigma_{(-3,3)\rightarrow (-2,2)}= 8 \pi \left(\frac{1}{77} a_6 +\frac{3}{11} a_4 - \frac{2}{7} a_0\right)^2 \\ \nonumber
\sigma_{(-3,3)\rightarrow (-1,1)}= 8 \pi \left(\frac{5}{154} a_6 +\frac{6}{154} a_4  - \frac{5}{14} a_2 + \frac{2}{7} a_0\right)^2  \\ \nonumber
\sigma_{(-3,3)\rightarrow (0,0)}= 16 \pi \left(\frac{5}{231} a_6 -\frac{9}{77} a_4 + \frac{5}{21} a_2 - \frac{1}{7} a_0\right)^2
\label{sig}
\end{eqnarray}
For each channel, the cross-sections are found within the Born approximation by projecting both the initial and the final spin states into the symmetric molecular potentials $S=(6,4,2,0)$ (with corresponding scattering lengths $a_S$) in which interactions are diagonal.

\begin{figure}
\centering
\includegraphics[width= 3.0 in]{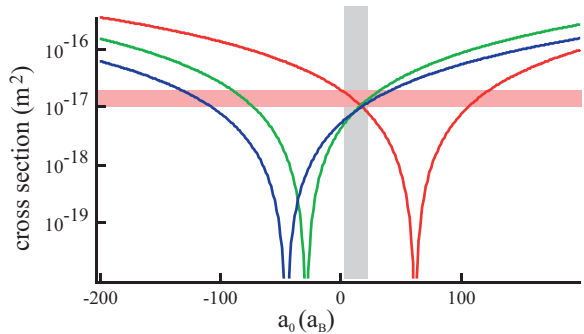}
\caption{\setlength{\baselineskip}{6pt} {\protect\scriptsize (Color online) Cross sections for processes (i) $\sigma_{(-3,3)\rightarrow (-2,2)}$ (red), (ii) $\sigma_{(-3,3)\rightarrow (-1,1)}$ (green), and (iii) $\sigma_{(-3,3)\rightarrow (0,0)}$ (blue) as a function of $a_0$. Their poles are located at +66, -29 and -48 $a_B$ respectively. Shaded area is the experimental outcome.
}}
\label{fitdata0}
\end{figure}

In chromium, all scattering lengths are now well established, but $a_0$ \cite{feshbachpfau,pasquiou2010}. We therefore can fit our experimental data using the known values of the scattering lengths $a_{6,4,2}$ given in these references. Because of the parabolic dependency of $\sigma_{(-3,3)\rightarrow (-2,2)}$ as a function of $a_0$, it is however not possible to unambiguously deduce the value $a_0$ from only $\sigma_{exp, (-3,3)\rightarrow (-2,2)}$. Fortunately, our experimental data also provide information for the channels of spin exchange (ii) and (iii). As shown in Fig. \ref{dynamics2}, we find that the spin dynamics rates for $m_s=-1$ and $m_s=0$ are comparable to the one for $m_s=-2$. Figure \ref{fitdata0} shows the experimental evaluation of the cross-section for each channel, as well as the predicted values as a function of $a_0$. As seen from the figure, large positive or negative values of $a_0$ are clearly excluded: only values of $a_0$ close to 0 lead to comparable rates for the three dynamics. We infer from our experimental data that $a_0=12^{+15}_{-10} a_B$.

As the data in Fig. \ref{dynamics2} show non-negligible fraction of atoms in other spin states than $m_s=\pm 3$, we compared the experimental data with a simulation including all possible collision channels, and not only the leading channels at short time described above. For a given $m$ state, the full dynamics reads:

\begin{eqnarray}
\frac{dn_{m}}{dt}&=& \bar v \sum_{m_1,m_2}\Big( \sigma_{(m_1,m_2) \rightarrow (m,m')} n_{m_1} n_{m_2}\nonumber\\
&&-  \sigma_{(m,m') \rightarrow (m_1,m_2)} n_{m} n_{m'}\Big)
\label{dyntot}
\end{eqnarray}

with $m'=m_1+m_2-m$.
Fitting the simulation to the experimental data for population dynamics during the first 50 ms in $m_s=-2,-1,0$ lead to $a_0=13^{+15}_{-10}a_{B},13.5^{+10.5}_{-13.5},1.5^{+7.5}_{-9.5}a_{B}$ respectively. The error bars correspond to a 66\% confidence interval. These values are all compatible with each other, and compatible with the result given by the simpler model. As absolute populations measurements in $m_s=0$ may be underestimated (due to a less efficient absorption process), we believe that there is a slight bias for the corresponding inferred value of $a_0$ towards the negative side. Therefore, we rely on the data for $m_s=-2,-1$ to make our final estimate, and obtain $a_0=13.5^{+11}_{-10.5}a_{B}$ with 95\% confidence.

We note that the good agreement between our theoretical treatment and the simple model described earlier confirms that most of the $m_s=-2$ population is created through channel (i), and that most of the $m_s=-1$ population is created through channel (ii). Interestingly, channel (i) does not depend on $a_2$. Therefore, having determined $a_0$ through the time evolution of $m_s=-2$, it is also possible to give an estimate of $a_2$ from the time evolution of $m_s=-1$. We infer a value of $a_2$ small and negative, in good agreement with \cite{feshbachpfau}, where the value of $a_2$ was deduced from the observation of one isolated Feshbach resonance.

\subsection{Conclusion}
We have experimentally studied the spin dynamics of a chromium BEC loaded into a double-well trap.  The well separation was sufficient to enable a spin preparation with opposite spin polarizations of the two atomic subsystems. We find that this spin configuration is metastable, with a lifetime largely exceeding the timescale associated with dipole-dipole interactions between clouds. This stability arises because a dynamical modification of the longitudinal magnetization under the influence of the exchange term of dipolar interactions would violate energy conservation, due to the non-negligible Ising dipolar term. The spin dynamics is therefore classically suppressed by the interplay between exchange and Ising interactions. This classical behavior is well accounted for by the very large effective spin realized within each well, in contrast to the situation studied earlier where at most pairs of atoms would undergo spin-exchange dipolar interactions \cite{3DSpinExchange}. In a second series of experiment, we started with the same prepared two ensembles of opposite spin polarizations, merged them and monitored the spin composition of the reunited system as a function of time. From the observed spin dynamics, we have obtained the first determination of the scattering length $a_0$ for collision in the S=0 channel. Our measurements yield a value of $a_0$ close to 0, but most probably positive. It is therefore likely that the low magnetic field spinor ground state is cyclic rather than polar \cite{spinorcr,shlyapnikov}.

 %In case the dimensionality of the gas is reduced to 1D, an interesting phase of condensed pairs of atoms is then possible at low temperature \cite{shlyapnikov}.

We thank Martin Robert de Saint Vincent for triggering our interest on the lateral displacement beam splitter. We thank Benoit Darqui\'e for his help with data analysis.

Acknowledgements: LPL is Unit\'e Mixte (UMR 7538) of CNRS and
of Universit\'e Paris 13, Sorbonne Paris Cit\'e. We acknowledge financial support from Conseil R\'{e}%
gional d'Ile-de-France under DIM Nano-K / IFRAF, and from Minist\`{e}re de
l'Enseignement Sup\'{e}rieur et de la Recherche within CPER Contract.

\end{document}